\documentclass{emulateapj}
\usepackage{url}
\usepackage{verbatim}
\usepackage{amsmath, amsthm, amssymb}

\newcommand{\etal}{{\it et al.}}
\newcommand{\arcm}{{$^\prime\,$}}

\begin{document}

\title{Constraining Source Redshift Distributions with Gravitational
  Lensing}
\shorttitle{Constraining Source Redshift Distributions}

\author{D. Wittman and W.~A. Dawson}
\affil{Physics Department, University of California, Davis,
  CA 95616}
\email{dwittman@physics.ucdavis.edu}


\begin{abstract} 
  We introduce a new method for constraining the redshift distribution
  of a set of galaxies, using weak gravitational lensing shear.
  Instead of using observed shears and redshifts to constrain
  cosmological parameters, we ask how well the shears around clusters
  can constrain the redshifts, assuming fixed cosmological parameters.
  This provides a check on photometric redshifts, independent of
  source spectral energy distribution properties and therefore free of
  confounding factors such as misidentification of spectral breaks.
  We find that $\sim 40$ massive ($\sigma_v=1200$ km s$^{-1}$) cluster
  lenses are sufficient to determine the fraction of sources in each
  of six coarse redshift bins to $\sim$11\%, given weak (20\%) priors
  on the masses of the highest-redshift lenses, tight (5\%) priors on
  the masses of the lowest-redshift lenses, and only modest (20-50\%)
  priors on calibration and evolution effects.  Additional massive
  lenses drive down uncertainties as $N_{\rm lens}^{-{1\over 2}}$, but
  the improvement slows as one is forced to use lenses further down
  the mass function.  Future large surveys contain enough clusters to
  reach 1\% precision in the bin fractions if the tight lens mass
  priors can be maintained for large samples of lenses.  In practice
  this will be difficult to achieve, but the method may be valuable as
  a complement to other more precise methods because it is based on
  different physics and therefore has different systematic errors.
\end{abstract}

\section{Introduction}
Photometric redshifts are of key importance to current and future
galaxy surveys.  In a typical application, galaxies might be binned
according to photometric redshift and then some analysis conducted on
these binned source sets.  Avoiding systematic error in this case
requires knowledge of the true redshift distribution of each
photometric redshift bin.  Ideally, photometric redshift methods
themselves produce reliable confidence intervals for this purpose, but
external verification is important for controlling and quantifying
systematic errors.  For deep surveys, direct verification with a
spectroscopic sample representative of the photometric sample is very
difficult, and this has led to the development of other methods such
as cross-correlating each photometric redshift bin with other
photometric redshift bins (Schneider \etal\ 2006, Erben \etal\ 2009)
or with spectroscopic redshift bins (Newman 2008, Matthews \& Newman
2010).  Here we present a new and independent method for
reconstructing source redshift distributions, using the shear from
weak gravitational lensing.

The shear induced by a given lens grows with source redshift in a
well-understoood way which depends on cosmographic parameters.  Other
authors (Jain \& Taylor 2003, Bernstein \& Jain 2004) have suggested
using this dependence to constrain the cosmographic parameters from
the observed shear-vs-redshift relation around identified lenses.
(Note that this is distinct from cosmic shear, for which the redshift
dependence also involves the growth of structure.)  In this paper we
reverse the question and ask how well the redshift distribution of a
set of galaxies can be constrained by shear measurements around lenses
at a range of redshifts, if the background cosmology is already well
determined (see Appendix~\ref{sec-cosmo} for a demonstration that
cosmological uncertainties are negligible here).  This may prove
useful for applications which take the background cosmology as given,
and it also provides an internal consistency check for surveys.

Although the shear around any given lens is rather weak and might be
expected to provide very little constraint, a very large survey such
as LSST (Ivezi\^{c} {\it et al.} 2008) will contain billions of
sources lensed by tens of thousands of galaxy clusters.  Assuming the
photometry is uniform over the sky, the entire dataset can be used to
constrain the source redshift distribution.  In this paper we use
Fisher matrix formalism to estimate an upper bound on the
effectiveness of this method, and we discuss challenges to
implementing the method on real data.  Because this method requires a
very large survey, we refer to LSST throughout.  We refer the reader
to the LSST Science Book (LSST Science Collaborations {\it et al.}
2009) for more details on LSST survey parameters.

\section{Basic idea and toy model}\label{sec-toy}

\begin{figure}
\begin{centering}
\includegraphics[scale=0.45]{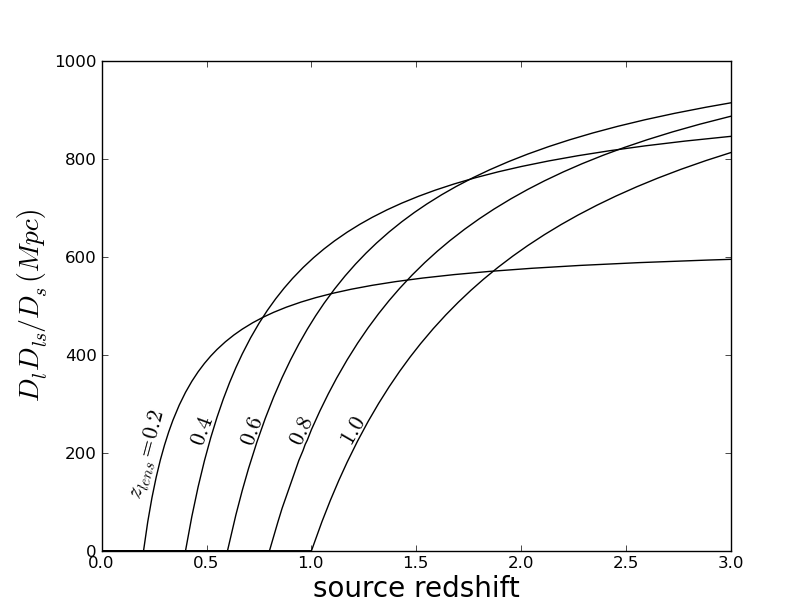}\\ \includegraphics[scale=0.45]{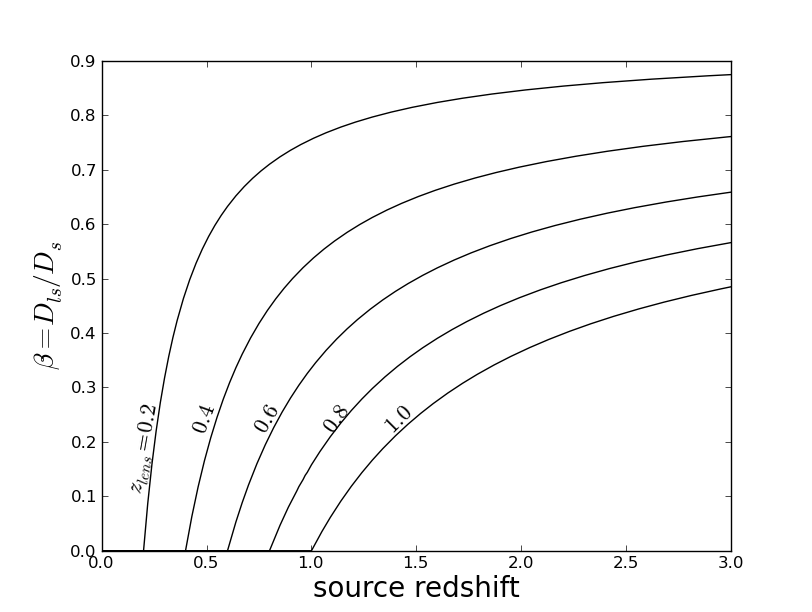}\\
\end{centering}
\caption{{\it Top:} The distance factor ${D_{ls} D_{l}\over D_s}$ as a
  function of source redshift, for various fixed lens redshifts.  The
  lens redshift for each curve can be read off at the point where the
  curve departs from zero. {\it Bottom:} as for the left panel, but
  plotting $\beta={D_{ls}\over D_s}$. \label{fig-dratio}}
\end{figure}

The tangential shear $\gamma$ at a projected distance $r$ around an
axisymmetric lens is:
\begin{equation}
\gamma(r) = {4\pi G\over c^2} {D_{ls}D_l\over
  D_s}[\bar{\Sigma}(<r)-\Sigma(r)]
\end{equation} where $D_{ls}$, $D_l$, and $D_s$
are respectively the angular diameter distances from lens to source,
observer to lens, and observer to source, $\Sigma(r)$ is the projected
surface mass density at $r$, and $\bar{\Sigma}(<r)$ is the mean
projected surface mass density within $r$ (Miralda-Escud\`{e} 1991).
Figure~\ref{fig-dratio} (top panel) shows the behavior of the distance
factor ${D_{ls} D_{s}\over D_s}$ as a function of source redshift, for
various fixed lens redshifts. This depends on the assumed cosmological
model; we assume a WMAP7 $\Lambda$CDM cosmology of $H_0 = 70.4$ km
s$^{-1}$ Mpc $^{-1}$, $\Omega_m = 0.272$, and $\Omega_\Lambda = 0.728$
throughout.  It is clear that very low-redshift ($z<0.2$) source
galaxies will have no measurable shear around any of the lenses in the
figure, moderate-redshift ($\sim 0.5$) sources will have measurable
shear only around the lower-redshift lenses, and high-redshift ($z>1$)
sources will have measurable shear around all lenses.  This is the
conceptual basis for constraining the redshift distribution of a
source set: one should be able to solve for the source redshift
distribution which, when passed through the appropriate response
curves, best matches the observed shears.  The method requires lenses
at a range of redshifts and, to obtain good constraints, some prior
knowledge of the lens masses.  To simplify the exposition we treat the
lens propertise as known in this section, and explore requirements on
prior knowledge of lens properties in the next section.

Individual galaxies as well as clusters of galaxies can serve as
lenses.  Typically, the shear one arcminute off axis might be of order
0.5 for clusters and 0.01 for galaxies.  Galaxies have near-isothermal
mass profiles (Gavazzi {\it et al.} 2007), in which $\Sigma(r)$ and
$\bar{\Sigma}(<r)$ scale as $r^{-1}$; specifically,
$\bar{\Sigma}(<r)-\Sigma(r) = {\sigma_v^2\over 2Gr}$ where $\sigma_v$
is the velocity dispersion.  Because this scales as $r^{-1}$, it
scales as $(\theta D_l)^{-1}$ in terms of projected {\it angular}
distance $\theta$, and $\gamma(\theta)$ can be written
$$\gamma(\theta) = {2\pi \over c^2} {D_{ls}\over D_s} {\sigma_v^2\over
  \theta}$$ Isothermal lenses thus have the convenient property that
$\gamma(\theta)$ can be separated into a part which depends {\it only}
on the cosmology and a part which depends {\it only} on the mass or
velocity dispersion of the lens.  Mass profiles which are not
scale-free cannot be written this way because the conversion from
physical to angular scales invokes the cosmological model through
$D_l$.  This convenience is not required for our argument, but will
substantially simplify the exposition here.  Although cluster mass
profiles depart from isothermality more than galaxy profiles do, for
this section we will assume isothermality to simplify our argument.

We can now write a very simple model for the shear of the $i$th source
galaxy due to the $j$th lens.  We henceforth refer to $\sigma_v$ as
$s$ to eliminate confusion with the symbol $\sigma$ which we will use
later to indicate uncertainties on measurements.  We also define
$\beta = {D_{ls}\over D_{s}}$ (taking $\beta$ to be zero where $z_s\le
z_l$); the $\beta$ curves for various lenses are shown in the bottom
panel of Figure~\ref{fig-dratio}.  We can then write
\begin{equation}
\gamma_{ij} = {2\pi \over c^2} \beta_{ij} {s_j^2\over \theta_{ij}}
\label{eqn-shearij}
\end{equation}
where $\theta_{ij}$ is the angular separation between lens $j$ and
source $i$, $\beta_{ij}$ is the distance ratio between the two, and
$s_j$ is the velocity dispersion of lens $j$.  

The redshift distribution of the source set can be parameterized in
many ways.  Here we adopt the simplest approach, uniform bins in
redshift.  With $N_{\rm bin}$ bins, we parametrize the distribution
with $f_1,f_2,...f_{N_{\rm bin}}$, where $f_m$ is the fraction of the
galaxies in the set which are actually in redshift bin $m$.  The
highest-redshift bin should extend to arbitrarily high redshift so
that none of the measured shear can come from an unmodeled part of the
distribution.\footnote{Because $\beta$ becomes fairly flat at high
  redshift, a very wide high-redshift bin is no more problematic than
  a modest-width low-redshift bin in terms of variation of $\beta$
  within each bin.}  Consequently, we do not require that
$\sum_m f_m=1$; any difference between unity and $\sum_m f_m$ can be
interpreted as due to sources at such low redshift that they are never
lensed.  (A derived parameter $f_0 \equiv 1-\sum_m f_m$ can be defined
which encodes the fraction of sources at ultralow redshifts, but is
not a parameter we solve for directly.)  For sources from an unknown
redshift distribution, the expected value of $\beta$ is simply $\sum_m
f_m \beta_{mj}$, where $\beta_{mj}$ is now the mean distance ratio
between lens $j$ and source redshift bin $m$.  Thus the expected shear
for a lens-source combination is
\begin{equation}
\langle \gamma_{ij}\rangle = {2\pi \over c^2} {s_j^2\over \theta_{ij}}\sum_m f_m \beta_{mj} 
\label{eqn-shearij2}
\end{equation}
In this section, we will assume that $s_j$ is perfectly known; we will
consider uncertainty in lens masses in \S\ref{sec-mass}.  

The observable is the total shear\footnote{Technically, the observable
  is the reduced shear $\frac{\gamma}{1-\kappa}$ where $\kappa$ is the
  convergence, but the argument still holds.  The effect of $\kappa$
  is to change the profile somewhat, which is not important for the
  toy model here.} on each source galaxy due to all lenses which, in
the weak lensing limit, is $\gamma_i = \sum_j \gamma_{ij}$.  However,
we must add shears componentwise because in the multiple-deflector
context there is no longer a single ``tangential'' component.  Thus
shear is usually written as a complex quantity:
$$\langle\gamma_{ij}\rangle = {2\pi \over c^2} {s_j^2\over
  \theta_{ij}}e^{i2\phi_{ij}} \sum_m f_m \beta_{mj} $$ where
$\phi_{ij}$ is the position angle of source $i$ with respect to lens
$j$, and the $i$ in the exponent represents $\sqrt{-1}$.  The total
expected shear on the source is then
$$\langle\gamma_i\rangle = \sum_j {2\pi \over c^2} {s_j^2\over \theta_{ij}}e^{i2\phi_{ij}}
\sum_m f_m \beta_{mj}.$$  

We could form a Fisher matrix with each two-component $\gamma_i$
contributing two observables; the derivatives of these observables
with respect to the model parameters $f_m$ are very straightforward.
However, this is computationally inefficient because most sources will
be far from any modeled lens (assuming we use massive clusters rather
than individual galaxies as the lenses) and thus will contribute
negligible information to the Fisher matrix; and of the remaining
sources, most are near only one lens so that their contribution can be
captured by a single (tangential) component.  Thus, at the expense of
very little information, we can efficiently group sources according to
their relevant lenses.

To do this, we group the observables in Equation~\ref{eqn-shearij2}
on one side and the model parameters on the other:
\begin{equation}
\langle\gamma_{ij}\theta_{ij}\rangle = {2\pi \over c^2} {s_j^2} \sum_m f_m \beta_{mj}.
\label{eqn-shearij3}
\end{equation}
The $e^{i2\phi_{ij}}$ factor no longer appears because with only one
lens involved the tangential component is unambiguous.  We can
collapse this down to a single observable for each lens by taking a
mean of sources relevant to that lens:
\begin{equation}
\Gamma_j \equiv \langle\gamma_{ij}\theta_{ij}\rangle_{i\in j} = {2\pi \over
  c^2} {s_j^2} \sum_m f_m \beta_{mj}\label{eqn-Gammadef}
\end{equation}
where ${i\in j}$ denotes averaging over all sources projected
close enough to lens $j$ to add non-negligible information to the
Fisher matrix.  Accordingly,
\begin{equation}
{\partial \Gamma_j \over \partial f_m} = {2\pi s_j^2\over c^2}
\beta_{mj} 
\end{equation}

To construct the Fisher matrix elements, we also need the variance of
$\Gamma_j$.  Because there is essentially no uncertainty on
$\theta_{ij}$, the variance of $\gamma_{ij}\theta_{ij}$ is
$\sigma_{i}^2\theta_{ij}^2$ where $\sigma_{i}$ is the uncertainty on
the measured shear of galaxy $i$.  Furthermore, $\sigma_{i}$ is
uncorrelated with position or lens, so we can approximate it as a
constant $\sigma_\gamma$ and pull it out of any sums.  Weighting
sources by inverse variance when computing $\Gamma_j$, we then find
that
\begin{equation}
\sigma_{\Gamma_j}^{-2} =  \sigma_\gamma^{-2}\sum_{i \in j} {1\over \theta_{ij}^2}\label{eqn-sigmaGamma}
\end{equation}
This sum primarily involves properties of the survey rather than the
lens.  We can rewrite it as the integral of the areal density of the
source set, $n$, over the ``footprint'' of the lens, that is, the area
over which the lens contributes non-negligible Fisher information:
$n\int \theta^{-2} dA$.  (Note that $n$ here is in units of galaxies
per steradian, because $\theta$ is in units of radians.  We will
convert to the more typically quoted units of arcmin$^{-2}$ at the end
of the calculation.)  We shall see that the value of this integral is
only weakly related to the properties of the lens.  Integrating from
some $\theta_{\rm min}$ to some $\theta_{\rm max}$ yields
\begin{eqnarray}
n\int \theta^{-2} dA& =& 2\pi n \int_{\theta_{\rm min}}^{\theta_{\rm max}}
\theta^{-1} d\theta\\
& = & 2\pi n \ln(\frac{\theta_{\rm max}}{\theta_{\rm min}})\label{eqn-ln}
\end{eqnarray}
Because the dependence is only logarithmic, we can make a rough
estimate of $\frac{\theta_{\rm max}}{\theta_{\rm min}}$ for a typical
lens and apply it to all lenses without much loss of precision.
Sources less than $\sim 1$\arcm\ from the lens center are typically
not used in weak lensing analyses because of the risk of contamination
by cluster members and/or errors on their shear measurements due to
background gradients from cluster members, so we adopt this value of
$\theta_{\rm min}$. At large enough $\theta$, shear from unrelated
structures will dominate the shear from the cluster.  Although
information about the cluster is still there, as a practical matter
the effort to recover it may not be justified given the logarithmic
dependence on $\theta_{\rm max}$.  The shear around a massive cluster
can clearly be observed out to at least $15$\arcm\ from the center
(e.g, Kling \etal\ 2005), so $\theta_{\rm max}=15$\arcm\ should be a
conservative value.  Thus Equation~\ref{eqn-ln} is approximately equal
to $17n$, and only weakly dependent on the adopted values of
$\theta_{\rm min}$ and $\theta_{\rm max}$.  We expect the dependence
on assumed shear profile to be weak as well, given that many
ground-based weak-lensing cluster studies have difficulty
discriminating between different forms of shear profiles (Wittman
2002).  Inserting this result into Equation~\ref{eqn-sigmaGamma}, we
find that
$$\sigma_{\Gamma_j}^2= \frac{\sigma_\gamma^2}{17n}.$$ 

Assuming a Gaussian likelihood model, we now have an $N_{\rm bin}$ by
$N_{\rm bin}$ Fisher matrix:
\begin{eqnarray}
\mathcal{F}_{mn} &=& \sum_j^{N_{\rm lens}} {1 \over \sigma_{\Gamma_j}^2}
{\partial \Gamma_j \over \partial f_m} {\partial \Gamma_j \over
  \partial f_n}\\
&=& \left({2\pi\over c^2}\right)^2 {17n\over\sigma_\gamma^2}\sum_j^{N_{\rm lens}}
s_j^4 \beta_{mj} \beta_{nj} \label{eqn-betas}\\
& \approx & 25 (\frac{n}{\rm arcmin^{-2}}) \sum_j^{N_{\rm lens}}
(\frac{s_j}{\rm 1000\ km\ s^{-1}})^4 \beta_{mj} \beta_{nj}  
\end{eqnarray}
where we have adopted $\sigma_\gamma = 0.2$, which represents the
irreducible ``shape noise'' due to the random intrinsic orientations
of sources.  Smaller, fainter, sources will have larger shear
uncertainties due to measurement uncertainties: $\sigma_i^2 =
\sigma_\gamma^2 + \sigma_{{\rm meas},i}^2$.  We account for this by
defining $n$ as the {\it effective} source density, that is, the
equivalent number of perfectly measured sources provided by the source
set: ${n} = \sigma_\gamma^2 \sum {1\over \sigma_\gamma^2 +\sigma_{{\rm
      meas},i}^2}$ where the sum is taken over a unit area of sky.
This is the basis on which surveys or source sets should be quoted;
see the explanation of Equation 14.7 in the LSST Science Book (LSST
Science Collaborations {\it et al.} 2009) for a more detailed
justification.

As a rough estimate of the potential of this method, we compute a
Fisher matrix forecast for an LSST-depth ($n=40$ arcmin$^{-2}$) survey
covering nineteen lenses with $\sigma_v=1200$ km s$^{-1}$ at redshifts
$z=\{0.05,0.1,0.15,...,0.95\}$, which roughly corresponds to using the
most massive lens in the sky at each redshift.  We model six redshift
bins: five covering $0<z \leq 1$ with equal spacing, plus a sixth bin
containing all galaxies at $z>1$.  The forecast uncertainties on $f_m$
are then $\{0.06,0.07,0.11,0.17,0.20,0.10\}$.  The same Fisher
information can be obtained using a larger number of less-massive
lenses, but going beyond a small number of well-studied lenses
introduces a new problem: lens parameters such as $\sigma_v$ will not
be known precisely for all these lenses, and we must examine the
effect of marginalizing over uncertainties in the lens properties.



\section{Marginalizing over lens mass}\label{sec-mass}

The two lens parameters in this model are redshift and mass (or
equivalently, velocity dispersion).  Spectroscopic redshifts are
available in the literature for thousands of clusters, particularly
for the most massive ones which contribute the most Fisher
information.  But with large imaging surveys now finding more clusters
than can be followed up with spectroscopic redshift confirmation,
photometric redshifts for the lenses must be considered.  Photometric
redshifts are much more precise for clusters than for individual
galaxies, even beyond the statistical effect of averaging many member
galaxies.  Because cluster members are preferentially early-type
galaxies with well-established spectral energy distributions, their
redshifts can be established with very high confidence.  Furthermore,
the accuracy of cluster photometric redshifts can be easily verified
by spectroscopy of a representative subsample of clusters, whereas
representativeness is very difficult to achieve in spectroscopic
subsamples of individual galaxies.  The LSST Science Book (LSST
Science Collaborations {\it et al} 2009) forecasts photometric
redshift uncertainties for modest (10$^{14} M_\odot$) clusters as a
function of redshift; their Figure 13.12 shows that the uncertainties
are 0.01 or less for all $z\le 1.1$ after a single visit (and for all
$z\le 1.9$ after ten years of the survey).  This fixes the $\beta$
factors to within 0.01, which is far more precise than the masses will
be known {\it a priori}.  We can therefore safely ignore uncertainties
in lens redshifts.  In the remainder of this section, we focus on
marginalizing over lens mass uncertainties.

We now regard the $s_j^2$ as nuisance parameters. We choose $s_j^2$
rather than $s_j$ simply for convenience: $s_j^2$ is conceptually
convenient because it is linearly proportional to the shear, and this
proportionality makes it fall out of the relevant Fisher matrix
elements.  From Equation~\ref{eqn-Gammadef} we have
\begin{equation}
{\partial \Gamma_j \over \partial (s_k^2)} = \begin{cases}
{2\pi \over c^2} \sum_m f_m \beta_{mj}  & \text{if $j=k$,}\\
0 & \text{otherwise.}
\end{cases}
\label{eqn-betasum}
\end{equation}
We now have a Fisher matrix which is $N_{\rm bin}+N_{\rm lens}$
elements square.  The lower-right $N_{\rm lens}\times N_{\rm lens}$
block is diagonal, because ${\partial \Gamma_j \over \partial (s_k^2)}
= 0$ where $j \ne k$.  These diagonal elements look like the square of
the first case in Equation~\ref{eqn-betasum}.  The upper-left $N_{\rm
  bin}\times N_{\rm bin}$ block contains only elements of the form in
Equation~\ref{eqn-betas}, and the other blocks (the $N_{\rm
  lens}\times N_{\rm bin}$ block in the upper right and its transpose
in the lower left) contain mixed terms.  Note, however, that many of
the elements in these latter blocks will be zero, corresponding to
lens/bin combinations for which the bin redshift is lower than the
lens redshift, and thus for which $\beta_{mj}=0$.  In the upper left
block, no elements will vanish unless {\it all} lenses are at a higher
redshift than the lowest-redshift bin.  Note that the $f_m$ now appear
in the Fisher matrix, so we must adopt fiducial values.  Fisher matrix
estimates should not be trusted far from the fiducial values; see
Albrecht {\it et al.} (2006) for a clear exposition of Fisher matrix
limitations.  We therefore adopt a uniform fiducial distribution for
now, and examine the dependence on the fiducial distribution at the
end of this section.

\begin{figure}
\includegraphics[scale=0.45]{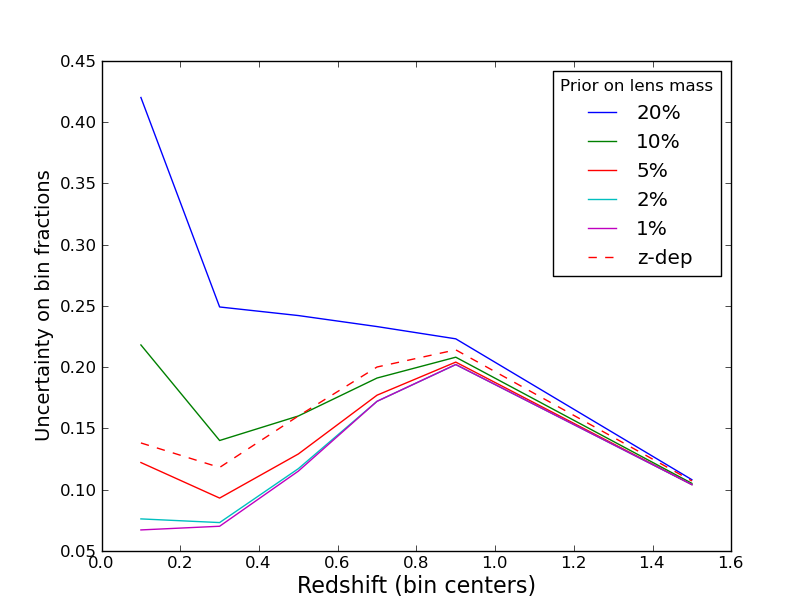}
\caption{Constraints on the bin fractions for various lens-mass priors
  in the simplified physical scenario of \S\ref{sec-toy} with just one
  massive lens at each redshift.  The redshift-varying prior allows
  for weaker lens-mass priors at higher lens redshifts where priors
  are both less necessary and more difficult to establish
  observationally; this particular form is $5\%(1+z_{\rm lens})^{2}$
  which brings the prior to 20\% at $z_{\rm lens}=1$. More lenses of
  the same mass will tighten the constraints as $N_{\rm
    lens}^{-{1\over 2}}$.
  \label{fig-masspriors}}
\end{figure}

Next, we need a prior on the lens masses.  With no prior, the
lowest-redshift lens would have no constraining power at all, as any
shear (or lack thereof) in the lowest-redshift bin could be explained
as a result of arbitrarily large (or small) lens mass.  Successively
higher-redshift lenses would have some constraining power, but not
much.  We therefore explore the importance of this prior by running
Fisher matrix forecasts with the lens set of the previous section, but
with priors of 1\%, 2\%, 5\%, 10\%, and 20\% on the lens masses
assuming the fiducial $s_j=1200$ km s$^{-1}$.  We implement this by
adding to the Fisher matrix the inverse variance representing a
Gaussian prior before inverting the Fisher matrix to obtain the
covariance matrix.  To implement the 1\% prior, for example, we add
$(0.01(1200^2))^{-2}$ to $\mathcal{F}_{7,7},
\mathcal{F}_{8,8},...\mathcal{F}_{25,25}$, which are the nineteen
elements representing the lens $s_j^2$.  Figure~\ref{fig-masspriors}
shows the results. For the low-redshift bins, obtaining constraints
close to those of the previous section requires a 1-2\% prior, but for
higher-redshift bins the required priors are more modest, 10-20\%.
Priors tighter than 1\% (not shown) do not yield any visible
improvement in this figure.

The most efficient observational approach may thus be to obtain tight
priors on low-redshift lenses using velocity dispersions, X-ray
analyses, external lensing analyses, etc, while obtaining only very
rough priors on the higher-redshift lenses.  We model this
scenario by testing a mass prior which is 5\% at $z_{\rm lens}=0$
but loosens as $(1+z_{\rm lens})^2$ (i.e., a 20\% prior at $z_{\rm
  lens}=1$); this is is shown as the dashed curve in
Figure~\ref{fig-masspriors}.  This yields source redshift constraints
as good as those obtained with a 5-10\% prior, but without much
difficult high-redshift work.  Indeed, optical richness alone can
provide a 20-30\% prior on the mass (Rykoff \etal\ 2011), thus
obviating the need for followup at the high-redshift, assuming there
is no redshift above which the optical richness estimator breaks down.

\begin{figure}
\includegraphics[scale=0.45]{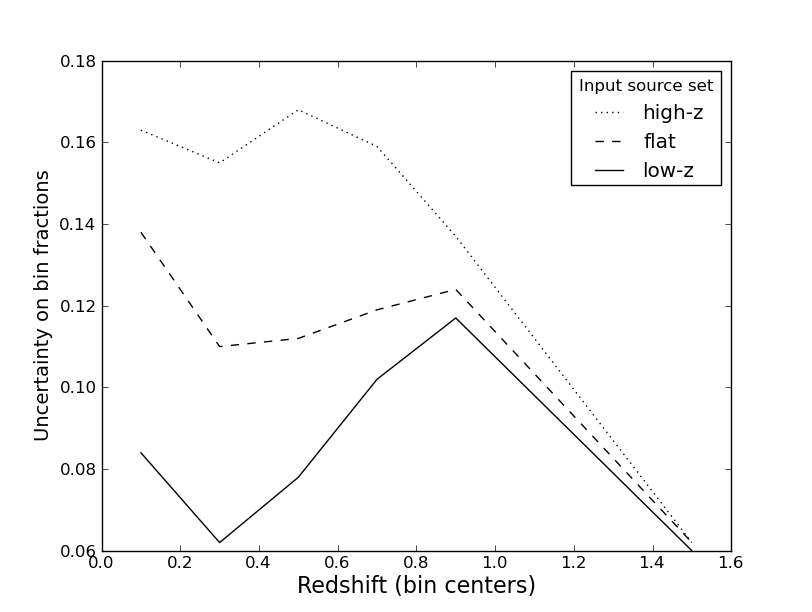}
\caption{Constraints on the bin fractions for low-redshift (lower
  curve), flat (middle curve), and high-redshift (upper curve) source
  sets.  Each source set is used with the redshift-varying lens-mass
  prior and 39 lenses described in the text.
  \label{fig-sourcedep}}
\end{figure}

To a reasonable approximation, each additional lens tightens the
constraints mostly on the 1-2 source redshift bins nearest the lens.
Therefore, we can tweak the lens sample to obtain roughly equal
constraints on all source redshift bins.  For example, we see from
Figure~\ref{fig-masspriors} that the maximum uncertainty for the
efficient redshift-varying mass prior is for source redshifts 0.7-0.9,
at a level of roughly twice the minimum uncertainty.  Tripling the
number of lenses at redshifts 0.5-0.95, for a total of 39 massive
lenses, yields roughly equal constraints on most source redshift bins,
averaging 0.11 (middle curve in Figure~\ref{fig-sourcedep}).  Reaching
yet lower levels of uncertainty requires more lenses overall.  For
example, reaching 0.01 uncertainty across the board will require
$11^2$ times more lenses than the previous scenario, or about 5000
total.  However, when so many lenses are considered, we cannot use
only very massive lenses. Therefore, the required number of lenses
will scale much more steeply than the inverse square of the desired
constraint.  The Fisher information corresponding to 5000 lenses with
$\sigma_v =1200$ km s$^{-1}$ could be accumulated with tens of
thousands of less-massive lenses, which is well within expectations
for current and future large imaging surveys.  Computationally, using
so many lenses should be feasible because most of the tens of
thousands of nuisance parameters can be marginalized over separately,
reflecting the nonoverlapping nature of most lenses.  However, the 1\%
goal will be difficult to achieve at low redshift due to the
difficulty of obtaining the 5-8\% lens-mass priors necessary to do so
well there.

These factors provide insight into why we do not propose using
individual galaxies as lenses.  For every $10^{15}$ $M_\odot$ cluster,
a survey would need one million $10^{12}$ $M_\odot$ galaxies to
provide the same Fisher information, all other factors being equal.
Although surveys such as LSST will have billions of galaxies, all
other factors are not equal.  The photometric redshift uncertainty on
each individual galaxy is much larger than for clusters, and
individual-galaxy lens photometric redshifts bring in exactly the
problems this method is designed to avoid.  These problems could be
avoided by using only early-type galaxies with well-behaved
photometric redshifts and mass priors from the fundamental plane, but
then far too few lenses would be available.  Furthermore, galaxy-scale
lenses do not dominate their local shear in a way which allows
parallelized modeling of designated patches of sky, and thus would
introduce additional computational challenges.

A foreseeable application for this method is to input source sets
corresponding to each of several photometric redshift bins, to
determine the outlier fraction and distribution in each.  Therefore,
we examine the dependence on the redshift distribution of the source
set, which we have assumed to be flat until now.
Figure~\ref{fig-sourcedep} shows the results using the
redshift-varying mass prior and the 39 massive lenses with three
different source sets: flat (as described above, middle curve), all in
the $z<0.2$ bin (lower curve), and all in the $0.8<z\le 1.0$ bin
(upper curve).  Characterization of outliers will be more precise for
the low-$z_{\rm phot}$ source set than for the high-$z_{\rm phot}$
source set, although this is partly due to the choice of lens-mass
prior which is tighter at lower redshift.

\section{Profile and calibration uncertainties\label{sec-profile}}

To this point we have assumed that prior knowledge of a cluster's
velocity dispersion\footnote{Throughout this section we refer to
  velocity dispersion for concreteness, but Sunyaev-Zel'dovich effect,
  X-ray, or optical richness measurements could equally well be used
  to set the prior.} translates perfectly into prior knowledge of its
integrated shear statistic, given a source redshift distribution.
However, the relationship between these two observables is sure to be
more complicated.  Clusters with the same velocity dispersion may have
different profiles, resulting in different integrated shear estimates.
If this is a purely stochastic effect, then the scatter in the
observable-observable relation weakens the prior.  But in principle,
the loss of information due to this scatter can be compensated by
analyzing more clusters.  Conceptually more worrisome would be a
redshift-dependent trend or an overall calibration error in this
relationship, which would cause systematic errors.

\begin{figure}
\includegraphics[scale=0.45]{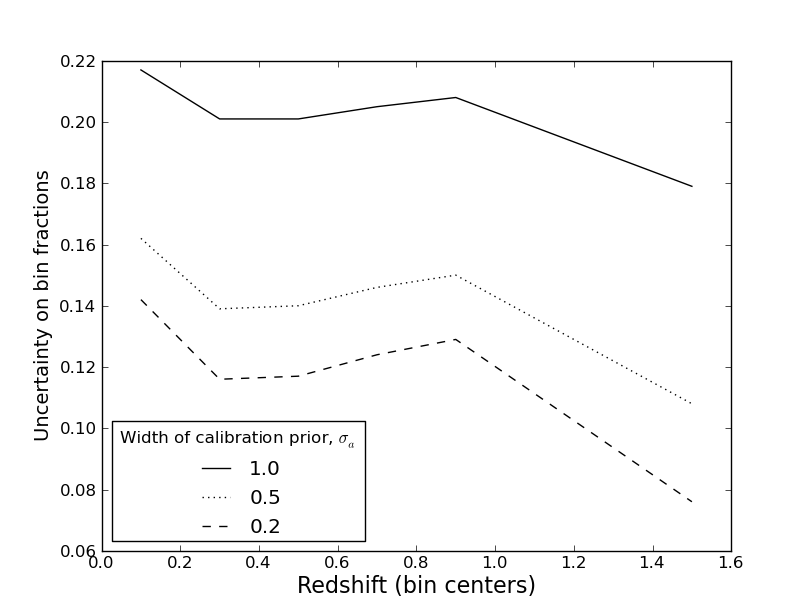}\\
\includegraphics[scale=0.45]{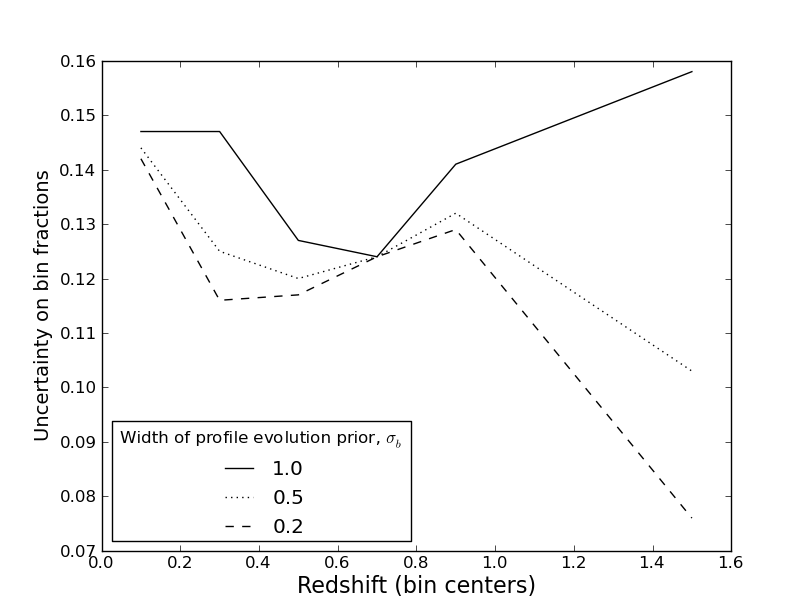}
\caption{Constraints resulting from a range of calibration (top panel)
  and evolution (bottom panel) priors.  For the top panel the
  evolution prior has been set to 20\% in all cases, and for the
  bottom panel the calibration prior has been set to 20\% in all
  cases.  Priors tighter than 20\% on either do not substantially
  improve the results: the bottom curve in each panel is very
  similar to the middle curve in Figure~\ref{fig-sourcedep}, where
  the priors were delta functions.
  \label{fig-evol}}
\end{figure}

We therefore insert nuisance parameters describing the uncertainties
in calibration and redshift evolution, and explore how the results
degrade as priors on these parameters are relaxed.  We allow
Equation~\ref{eqn-Gammadef} to be multiplied by a factor $a+bz_j$,
where $a$ is a nuisance parameter for the calibration, $b$ is a
nuisance parameter for the redshift evolution, and $z_j$ is the
redshift of the $j$th lens, and repeat the analysis of the previous
section with the flat source distribution and the redshift-dependent
lens mass prior.  Figure~\ref{fig-evol} shows the reconstruction
precision for a range of calibration and evolution priors, with a
fiducial model of $a=1$ and $b=0$.  Priors of 20\% on each are
sufficient to preserve performance nearly identical to that of the
previous section in which calibration and evolution were not factors;
the bottom curve in each panel is very similar to the middle curve in
Figure~\ref{fig-sourcedep}.

We stress that the choice of an isothermal profile as a calculation
aid here is not fundamental to the method.  In practice, surveys would
measure the average cluster profile and construct an optimal
integrated shear estimator before commencing the redshift analysis.
The results of this section indicate that profile calibration and
evolution uncertainties at the 20-50\% level do not substantially
degrade the performance of this method.

\section{Discussion\label{sec-detailedmodel}}

We have outlined a new method which uses gravitational lensing by
clusters and knowledge of the cosmology to constrain source redshift
distributions.  Although this is an unorthodox use of the
lensing-redshift information, it has some strengths and some relevant
applications.  The method's strengths are:
\begin{itemize}
\item it is independent of other methods.  It relies on no assumptions
  about galaxy spectral properties or biasing.

\item it is applicable to any set of galaxies.  One can determine the
  redshift distribution of a set of galaxies chosen in any number of
  ways, such as photometric redshift cut, pure color selection, or
  even X-ray or radio selection.  Multiwavelength observations are not
  necessary as they are for photometric redshifts.

\item in the context of a large imaging survey, it requires little
  additional data.  Where lens spectroscopic redshifts are not
  available, photometric lens redshifts will do, and optical richness
  provides a sufficient lens-mass prior for the high-redshift lenses.
  However, pinning down the low-redshift end of the source
  distribution will require strong mass priors on low-redshift lenses.

\end{itemize}

Our Fisher matrix forecast involves some approximations and caveats:
\begin{itemize}

\item Lens morphology: we have assumed axisymmetric isothermal lenses.
  Although we have argued that other profiles would not yield
  substantially different forecasts, a precise experiment would
  include some sort of marginalization over profiles.  This
  marginalization might include allowances for non-axisymmetry.

\item Foreground/background structure: we have assumed that shear from
  foreground/background structure is negligible in the ``footprint''
  of each lens, and totally dominant outside.  In reality, unmodeled
  shear will be present in the footprint as well, and contribute to
  slightly looser constraints.  However, unmodeled shear is not an
  irreducible source of noise.  One can choose to model previously
  unmodeled structure and thus add information to the Fisher matrix.
  This entails a cost-benefit analysis rather than a hard limit.

\item Magnification: the observed source population behind lenses will
  be slightly different than in other locations, due to lensing
  magnification.  If unmodeled, this may result in a nontrivial
  systematic error. However, because the lenses are being modeled, the
  model magnification at any given point provides an easily available
  basis for correction.  Deep fields which establish the source counts
  fainter than the main survey flux limit will be very useful for this
  purpose.  As with the foreground/background structure, it may be
  possible to add information to the Fisher matrix with improved
  modeling of the magnification.

\item Cluster contamination: the areas around clusters differ from the
  general area of a survey in another respect, the presence of cluster
  members.  Care must be taken so that the source set whose shear is
  being measured is not contaminated by these members.

\item Cosmology dependence of $\beta$: Appendix~\ref{sec-cosmo} shows
  that marginalizing over uncertainty in cosmographic parameters will
  not degrade the results presented here.  But using the cosmology as
  input here implies that the redshift estimates from this method
  should not feed in to some other method which estimates the
  cosmology unless the covariances are properly treated.

\item Wide source redshift bins: given a lens redshift, $\beta$ varies
  over the width of a source redshift bin.  Accurately modeling this
  effect will require some assumptions about the source redshift
  distribution inside the bins.  It may be possible to find a
  parameterization which reduces this difficulty.

\end{itemize}

The source redshift distributions obtained with this method will be
most useful in applications where the background cosmology would have
been assumed anyway, such as galaxy evolution, and especially in
applications which are very sensitive to photometric redshift
outliers.  For example, measurements of the bright end of the galaxy
luminosity function at high photometric redshift are extremely
sensitive to catastrophic outliers, as a small fraction of
low-redshift galaxies mistakenly put at high redshift results in many
more ``high-luminosity'' galaxies at that redshift.  Surveys can use
this method to independently derive the full redshift distribution of
the galaxies in each photometric redshift bin, thus characterizing the
number and distribution of outliers.  Indeed, the use of
multiwavelength photometry to infer both source redshift {\it and}
source spectral/photometric properties can be problematic, and this
method offers a possible workaround.

Other methods such as cross-correlation with a spectroscopic sample
(Newman 2008, Matthews \& Newman 2010) should be able to achieve 1\%
accuracy without assuming a cosmology or requiring expensive lens mass
priors.  This method may therefore not be competitive statistically,
but as it is based on different physics it may offer value in having
different systematic errors.  Redshift distributions derived using
this method can and should be used in consistency checks.  Given the
best-fit cosmology from weak lensing tomography, this method should
produce source redshift distributions which are consistent with those
used in the weak lensing tomography.  A prime application for this
method may therefore be as an internal consistency check for surveys.

\acknowledgments

We thank J.~A. Tyson for suggesting that we pursue the idea of using
lensing to constrain source redshift distributions, and the anonymous
referee for suggestions which led to improvements in the paper.

\appendix

\section{Dependence on Cosmological Parameters}\label{sec-cosmo}

This appendix justifies the assumption in the main body of the paper
that uncertainties in the distance ratios contribute negligible
uncertainty to the parameters of interest, the fraction of sources in
each redshift bin.  Uncertainty in the distance ratios comes from
uncertainty in cosmographic parameters, including Hubble's constant
$H_0$ and the energy densities in matter ($\Omega_m$), curvature
($\Omega_k$), and in the cosmological constant ($\Omega_\Lambda$).  Of
course, more general dark energy models can introduce further
parameters, but the cosmological constant has a greater effect on
distances than a dark energy model with equation of state parameter $w>-1$.
Therefore, if the assumption is justified for a cosmological constant
model, it is justified for models with $w>-1$.

Distances depend linearly on Hubble's constant $H_0$, but $\beta =
{D_{ls}\over D_{s}}$ is a {\it ratio} of distances.  Therefore $\beta$
does not depend on $H_0$ and uncertainty in $H_0$ is irrelevant.

The remaining parameters are more difficult to treat analytically, so
we make a numerical estimate.  The derivatives of $\beta$ with respect
to $\Omega_m$ and $\Omega_\Lambda$ (with
$\Omega_k=1-\Omega_m-\Omega_\Lambda$) depend on the lens and source
redshifts, but are typically of order 0.02.  The worst case is for
uncertainty in $\Omega_m$ when considering high-redshift sources and
lenses, where ${\partial \beta\over \partial \Omega_m} \approx 0.15$.
We take a conservative approach by setting ${\partial \beta\over
  \partial \Omega_m} = 0.15$ for all sources and lenses, so that the
partial derivative of Equation~\ref{eqn-Gammadef} with respect to
$\Omega_m$ is 0.15 times ${2\pi\over c^2}s_j^2$. We include $\Omega_m$
as a parameter in the Fisher matrix using these partial derivatives.


We rerun the forecast with a Gaussian prior of $\sigma=0.016$ on
$\Omega_m$, which corresponds to the current WMAP7+BAO+H0 uncertainty
for the ``OLCDM'' model in which curvature is allowed to
vary.\footnote{\url{http://lambda.gsfc.nasa.gov/product/map/dr4/params/olcdm\_sz\_lens\_wmap7\_bao\_h0.cfm}}
If the results were plotted on Figure~\ref{fig-masspriors} or
Figure~\ref{fig-sourcedep}, they would be nearly indistinguishable
from the existing plots.  The biggest difference would be the point in
the lower right corner of Figure~\ref{fig-sourcedep} increasing from
0.052 to 0.055.  The effect of varying $\Omega_\Lambda$ is even
smaller because $\beta$ is less sensitive to it.


\begin{thebibliography}{}

\bibitem[Albrecht at al.(2006)]{DETF} Albrecht, A., Bernstein, G.,
  Cahn, R., et al.\ 2006, arXiv:astro-ph/0609591

\bibitem[Bernstein \& Jain(2004)]{2004ApJ...600...17B} Bernstein, G.,
  \& Jain, B.\ 2004, \apj, 600, 17

\bibitem[Erben et al.(2009)]{2009A&A...493.1197E} Erben, T.,
  Hildebrandt, H., Lerchster, M., et al.\ 2009, \aap, 493, 1197

\bibitem[Gavazzi et al.(2007)]{2007ApJ...667..176G} Gavazzi, R., Treu, T., 
Rhodes, J.~D., et al.\ 2007, \apj, 667, 176 

\bibitem[Ivezic et al.(2008)]{2008arXiv0805.2366I} Ivezic, Z., Tyson, 
J.~A., Acosta, E., et al.\ 2008, arXiv:0805.2366 

\bibitem[Jain \& Taylor(2003)]{2003PhRvL..91n1302J} Jain, B., \&
  Taylor, A.\ 2003, Physical Review Letters, 91, 141302

\bibitem[Kling et al.(2005)]{2005ApJ...625..643K} Kling, T.~P., 
Dell'Antonio, I., Wittman, D., \& Tyson, J.~A.\ 2005, \apj, 625, 643 

\bibitem[LSST Science Collaborations et al.(2009)]{2009arXiv0912.0201L} 
LSST Science Collaborations, Abell, P.~A., Allison, J., et al.\ 2009, 
arXiv:0912.0201 

\bibitem[Matthews \& Newman(2010)]{2010ApJ...721..456M} Matthews,
  D.~J., \& Newman, J.~A.\ 2010, \apj, 721, 456

\bibitem[Miralda-Escude(1991)]{1991ApJ...370....1M} Miralda-Escude, J.\ 
1991, \apj, 370, 1 

\bibitem[Newman(2008)]{2008ApJ...684...88N} Newman, J.~A.\ 2008, \apj,
  684, 88

\bibitem[Rykoff et al.(2011)]{2011arXiv1104.2089R} Rykoff, E.~S., Koester, 
B.~P., Rozo, E., et al.\ 2011, arXiv:1104.2089 

\bibitem[Schneider et al.(2006)]{2006ApJ...651...14S} Schneider, M., Knox, 
L., Zhan, H., \& Connolly, A.\ 2006, \apj, 651, 14 

\bibitem[Wittman(2002)]{2002LNP...608...55W} Wittman, D.\ 2002, 
Gravitational Lensing: An Astrophysical Tool, 608, 55 

\end{thebibliography}
\end{document}